\documentclass[10pt,conference]{IEEEtran}

\usepackage{graphicx}
\usepackage{booktabs}
\usepackage{amsmath,amssymb}
\usepackage{url}
\usepackage{hyperref}
\usepackage{xspace}
\usepackage{listings}
\usepackage{multirow}
\usepackage{microtype}
\usepackage{enumitem}
\usepackage{balance}
\usepackage{array}
\usepackage{placeins}



\newcommand{\codebleu}{\textsc{CodeBLEU}\xspace}
\newcommand{\compileatk}{\texttt{compile@k}\xspace}
\newcommand{\passatk}{\texttt{pass@k}\xspace}

\title{LLMs as Idiomatic Decompilers: Recovering High-Level Code from x86-64 Assembly for Dart}

\author{
\IEEEauthorblockN{Raafat Abualazm\IEEEauthorrefmark{1} and Ayman Abo Elhassan\IEEEauthorrefmark{1}}
\IEEEauthorblockA{\IEEEauthorrefmark{1}Faculty of Engineering, Cairo University, Giza, Egypt\\
Email: raafat.202210476@eng-st.cu.edu.eg, ayman.abo.elmaaty@eng.cu.edu.eg}
}

\begin{document}
\maketitle

\begingroup
\footnotesize
\noindent\textit{\copyright~2026 IEEE. Personal use of this material is permitted. Permission from IEEE must be obtained for all other uses, in any current or future media, including reprinting/republishing this material for advertising or promotional purposes, creating new collective works, for resale or redistribution to servers or lists, or reuse of any copyrighted component of this work in other works.}
\par\endgroup
\vspace{0.4em}

\begin{abstract}
Translating machine code into human-readable high-level languages is an open research problem in reverse engineering. Despite recent advancements in LLM-based decompilation to C, \textbf{modern} languages like Dart and Swift are unexplored. In this paper, we study the use of small specialized LLMs as an idiomatic decompiler for such languages. Additionally, we investigate the augmentation of training data using synthetic same-language examples, and compare it against adding human-written examples using related-language (Swift $\rightarrow$ Dart). We apply \codebleu{} to evaluate the decompiled code readability and \compileatk{} to measure the syntax correctness. Our experimental results show that on a \textbf{73-function} Dart test dataset (representing diverse complexity levels), our 4B specialized model achieves 71.3 \codebleu{} (95\% CI 65.5--77.1), approximately comparable to a $\sim$480B code model (73.1; 67.4--78.8). On a \textbf{subset of 34 natural Dart functions}, it reaches \compileatk{}$_{5} = 79.4\%$ (Wilson 95\% CI 63.2--89.7), vs.\ 64.7\% (47.9--78.5) for the base model; the difference is suggestive but not statistically significant at 0.05. Our results indicate that adding Swift training data helps at 8B but not at 4B, suggesting a capacity threshold for effective cross-lingual transfer. Our experimental results show that small specialized models can generate readable, idiomatic Dart with meaningful identifiers while using minimal compute.
\end{abstract}


\section{Introduction}
Traditional decompilers like Hex-Rays IDA Pro and Ghidra produce functionally correct but hard-to-read code, with generic variable names (e.g., \texttt{v1}, \texttt{v2}) and a lack of semantic information~\cite{dire2019}. Recent machine learning methods have shown promising results in recovering more idiomatic code, primarily targeting C. Early efforts used RNNs~\cite{katz2018}, followed by Transformer-based models: SLaDe (200M parameters)~\cite{slade2024}, Nova (1B)~\cite{nova2023}, and LLM4Decompile (up to 33B)~\cite{llm4decompile2024}, achieving better improvements in improving decompiled code readability and re-executability.

Dart and Swift represent significant portions of modern software development, particularly in mobile and web applications. According to GitHub's language statistics, these languages power millions of applications and are growing rapidly in adoption. Their object-oriented paradigms, null-safety features, and extensive abstraction layers make them particularly challenging targets for decompilation, yet increasingly important for reverse engineering efforts.

Hosseini and Dolan-Gavitt showed that it is possible to decompile code into multiple languages (Go, Fortran, OCaml) without using language-specific tools~\cite{hosseini2022}. However, contemporary languages such as Dart and Swift remain unexamined due to their object-oriented paradigms and extensive abstractions.

Recent work has focused on making decompiler output easier to read. DeGPT~\cite{degpt2024} applies LLM refinement over classical decompiler output to reduce cognitive load, while Idioms~\cite{idioms2025} couples code and type prediction to improve exact-match on GNU coreutils. As well as the LLM4Decompile-Ref model that enhances upon Ghidra output These approaches primarily focus on C and demonstrate the limitations of existing methods when applied to modern languages with richer type systems and idioms.

We investigate small capacity-constrained LLMs, fine-tuned on targeted data, ability to produce idiomatic Dart/Swift from x86-64 assembly. Our central research question under limited target-language data is: Is it better to add more synthetic target-language pairs, or to add training data from a closely related language (Swift) with similar idioms and domain? To address this, we compare a Dart-only dataset (natural + synthetic) against a Dart+Swift dataset, applying a comprehensive evaluation protocol using \codebleu{} for code readability and \compileatk{} for syntax validation.

\textbf{Contributions.}
\begin{itemize}[leftmargin=*]
\item First end-to-end neural decompilation study from x86-64 assembly to Dart/Swift with idiom-aware evaluation.
\item A direct comparison of synthetic augmentation vs.\ adding a close language under low-data constraints.
\item A comprehensive evaluation protocol: \codebleu{} + \compileatk{} with uncertainty quantification on all metrics.
\item Practical training approach enabling 4B models to approach much larger code LLMs on Dart decompilation tasks.
\end{itemize}

\section{Related Work}

\subsection{Neural Decompilation for C}
The progression from RNN-based decompilers~\cite{katz2018} to Transformer-based systems has steadily improved both readability and re-executability of recovered C code. SLaDe~\cite{slade2024} adapted BART for assembly$\rightarrow$C, emphasising robustness to compiler optimizations. Nova~\cite{nova2023} leveraged StarCoder initialisation and code-pretrained priors, achieving significant improvements on standard benchmarks. LLM4Decompile~\cite{llm4decompile2024} scales up to 33B parameters and popularises evaluation grounded in execution success, reaching over 60\% re-executability on certain C benchmarks. 

These C-focused studies share common limitations: they primarily evaluate on simple algorithmic functions, struggle with optimized code, and produce outputs that lack domain-specific idioms. Our work addresses these gaps by targeting modern languages with richer abstractions and evaluating on more diverse, real-world functions.

\subsection{Language-Agnostic and Cross-Language Decompilation}
Beyond C,  Hosseini and Dolan-Gavitt~\cite{hosseini2022} explore retargetable decompilation (e.g., Go, Fortran, OCaml) by treating source and assembly as plain text, suggesting that multilingual decompilers are feasible without bespoke language front-ends. However, their evaluation was limited to simple programs and didn't address the challenge of language-specific idioms. Our design probes the data axis of that question: when the target language does not have large dataset pool (low-resource) (Dart), does mixing in a closely related high-level language (Swift) help a small model, or does it dilute the LLM's target signal? Complementary studies on multilingual code modelling and cross-lingual transfer show that transferability depends on both source/target pairs and model scale~\cite{multilingual2023,crosslingual2025}.

\subsection{Evaluation Metrics in Decompilation}
Previous decompilation works~\cite{slade2024,nova2023,llm4decompile2024} primarily relied on execution-based metrics like \passatk{} or simple token overlap measures. However, BLEU-like metrics miss structure; \codebleu{}~\cite{codebleu2021} augments n-gram matches with Abstract Syntax Tree (AST) and data-flow components to better capture semantic similarity in code. Still, compilation and execution are different notions of success. Hence, \compileatk{} results provide a measurable reference to indicate the output code's syntax validation. We therefore report \codebleu{} under a fixed decoding policy and separate executability via \compileatk{}, reserving semantic \passatk{} for future work when comprehensive unit tests become available.

\section{Dataset Construction}

We developed two main training datasets that include both natural and synthetic examples. The natural examples consist of 246 Dart functions compiled from RosettaCode projects, representing diverse algorithmic patterns. For synthetic augmentation, we employed multiple LLMs (GPT, Claude, DeepSeek and Qwen families) to generate diverse coding styles across different complexity levels, with all outputs validated for syntactic correctness through compilation. The synthesized data underwent rigorous quality control to ensure code correctness.

\textbf{Assembly Generation and Compilation:}
We used different compilation approaches for each language to balance training effectiveness with code preservation:
\begin{itemize}[leftmargin=*]
\item \textbf{Dart}: Compiled using \texttt{dart compile aot-snapshot}, which applies Dart's standard ahead-of-time (AOT) optimization including type propagation, inlining, and other production-level optimizations. We disabled tree shaking using \texttt{@pragma('vm:entry-point')} annotations to preserve function boundaries.
\item \textbf{Swift}: Compiled with optimization level \texttt{-O0} to preserve program structure and prevent aggressive optimizations that might obscure the source-to-assembly mapping during initial model training.
\end{itemize}

This compilation strategy means our Dart test evaluations reflect real-world optimized binaries, while the Swift training data (used only in cross-lingual experiments) represents unoptimized code. The optimization level mismatch between languages is discussed as a confounding factor in Section~\ref{sec:limitations}.

\textbf{Training Dataset Composition:}
\begin{itemize}[leftmargin=*]
\item \textbf{Dart-only dataset:} 1{,}194 examples (246 natural from RosettaCode, 948 synthetic) of assembly$\rightarrow$Dart pairs.
\item \textbf{Dart+Swift dataset:} 1{,}000 examples (246 Dart, 754 Swift) to evaluate cross-lingual training effects.
\end{itemize}

An important aspect of our synthetic augmentation is the inclusion of ``thinking tokens'' from DeepSeek-R1. These tokens provide chain-of-thought traces that help the model understand the reasoning process behind code transformations during training, improving the model's ability to recover semantic intent from assembly code.

The test set consisted of \textbf{73 held-out Dart functions} each with verified reference implementations. The 34 natural functions mentioned in the abstract are separate to these 73 functions, specifically those derived from real-world codebases rather than synthetic generation.

\section{Models and Training}

\subsection{Base Models}
We selected two state-of-the-art base models with complementary strengths. \textbf{Qwen3-4B-Thinking-2507} features 4.0B parameters with Grouped Query Attention and SwiGLU activation, supporting a 32k context window via strong-to-weak distillation~\cite{qwen3}. \textbf{DeepSeek-R1-0528-Qwen3-8B}, with 8B parameters distilled from DeepSeek-R1 to the original Qwen3-8B, was designed for multi-step reasoning tasks~\cite{deepseekr1}. We fine-tuned both models on our datasets to compare capacity effects.

\subsection{Fine-tuning Configuration}
We employed parameter-efficient fine-tuning using LoRA with DoRA enhancement~\cite{dora2024}, targeting all attention and FFN projections for maximum expressiveness. The configuration used rank $r{=}32$, $\alpha{=}32$, and dropout of 0.09 to prevent overfitting. Training utilized AdamW optimizer (paged, 8-bit) with a learning rate of $2\times 10^{-5}$, batch size of 1, and gradient accumulation of 4 steps. We used cross-entropy loss with label smoothing (0.1) to improve generalization. Training was conducted on a single NVIDIA H200 GPU, requiring approximately 1.5 hours for 4B models and 3 hours for 8B models.

\section{Evaluation Methodology}

\subsection{Syntactic/Semantic Similarity: \codebleu}
We employ \codebleu{}~\cite{codebleu2021}, which adds Abstract Syntax Tree (AST) and data-flow components to BLEU to better capture semantic similarity. Unlike traditional BLEU, \codebleu{} analyzes code structure through ASTs and tracks variable usage patterns, providing a more accurate measure of functional equivalence. To ensure fair comparison across all models, we maintain decoding parity: temperature $= 0.2$ (balancing creativity and consistency), top-$p = 0.99$ (nucleus sampling), and beam size $= 1$ (greedy decoding for reproducibility).

\subsection{Executability: \compileatk}
We evaluate \compileatk{}: the percentage of test examples for which at least one of the top-$k$ decoded hypotheses successfully compiles with the Dart compiler. For each input, we generate $k$ attempts with consistent parameters. We use $k{=}5$ (three attempts with beam $= 1$, two with beam $= 2$) and also report \compileatk$_1$ for single-attempt performance. Compilation indicates syntactic/type validity—not semantic equivalence—so we explicitly avoid equating it with functional correctness. We reserve \passatk{} (unit-test success) for future work.

\subsection{Uncertainty Quantification and Statistical Tests}
To provide rigorous evaluation, we report 95\% confidence intervals (CIs) for all metrics. For proportions (like compilation rates), we use Wilson intervals which provide better coverage for small samples. For means (like \codebleu{} scores), we use normal CIs computed as mean $\pm 1.96 \cdot \mathrm{SD}/\sqrt{n}$. Additionally, we perform two-proportion $z$-tests for compilation rate comparisons to assess statistical significance.

\begin{figure}[ht]
    \centering
    \includegraphics[width=\columnwidth]{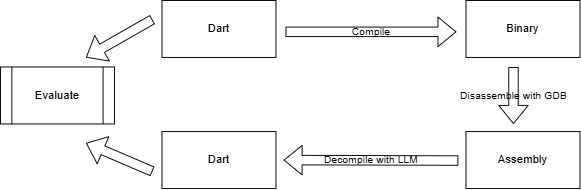}
    \caption{Decompilation pipeline: Dart/Swift source is compiled to binary, disassembled to x86-64 assembly, then decompiled by our fine-tuned models. We evaluate using \codebleu{} (code similarity) and \compileatk{} (syntactic correctness).}
    \label{fig:decompilation_pipeline}
\end{figure}

\section{Results}

\subsection{CodeBLEU Performance (Dart, $n{=}73$)}

To illustrate the qualitative differences in model outputs, we present example decompilations from a single test case. The following snippets show how the base model and our specialized models handle string interpolation and array operations from x86-64 assembly:

These examples demonstrate how specialized training affects decompilation quality. The base model produces compact but semantically incomplete code, while Decompiler-v1 generates more verbose output with detailed comments explaining assembly-level operations. Decompiler-v2 shows intermediate complexity with helper function abstractions. While all three outputs compile successfully, they differ significantly in idiomaticity and semantic fidelity to the original assembly.

Table~\ref{tab:codebleu} presents per-function \codebleu{} scores across model families under our fixed decoding policy. Our 4B Dart-specialized model (Decompiler-v1) achieves 71.3 \codebleu{}, representing a 5.2-point improvement over the base Qwen3-4B model. This performance is particularly noteworthy as it approaches the 73.1 score of Qwen3-Coder-Plus, a model with approximately 120× more parameters. The confidence intervals reveal consistent performance, with our specialized model showing lower variance (SD=25.4) compared to larger models.
\FloatBarrier
\begin{table}[!htbp]
\centering
\caption{\codebleu{} on the 73-item Dart test set (diverse complexity). CIs computed as mean $\pm 1.96 \cdot \mathrm{SD}/\sqrt{73}$; decoding: $t{=}0.2$, $p{=}0.99$, beam $=1$.}
\label{tab:codebleu}
\begin{tabular}{@{}lcccc@{}}
\toprule
\textbf{Model} & \textbf{Params} & \textbf{Mean} & \textbf{SD} & \textbf{95\% CI} \\
\midrule
Qwen3-4B (base) & 4B & 66.1 & 25.7 & 60.20--72.00 \\
Decompiler-v1 (4B Dart) & 4B & 71.3 & 25.4 & 65.47--77.13 \\
Decompiler-v2 (4B +Swift) & 4B & 69.2 & 20.1 & 64.59--73.81 \\
DeepSeek-8B (base) & 8B & 56.4 & 28.9 & 49.76--63.04 \\
Decompiler-v3 (8B Dart) & 8B & 59.5 & 30.6 & 52.48--66.52 \\
Decompiler-v4 (8B +Swift) & 8B & 68.2 & 23.9 & 62.72--73.69 \\
Qwen3-Coder-Plus ($\sim$480B) & $\sim$480B & 73.1 & 24.8 & 67.41--78.79 \\
Qwen3-Plus ($\sim$235B) & $\sim$235B & 60.3 & 31.4 & 53.10--67.50 \\
Qwen3-Max ($\sim$1T) & $\sim$1T & 77.4 & 19.4 & 72.95--81.85 \\
DeepSeek-Reasoner ($\sim$685B) & $\sim$685B & 74.6 & 24.7 & 68.94--80.27 \\
\bottomrule
\end{tabular}
\end{table}

Interestingly, the 4B model benefits more from Dart-only training (v1: 71.3) than from mixed Dart+Swift training (v2: 69.2), while the opposite pattern emerges at 8B scale, where Decompiler-v4 (8B +Swift) significantly outperforms v3 (8B Dart) by 8.7 points.

\subsection{Executability: \compileatk{} (Dart Natural Functions, $n{=}34$)}
Table~\ref{tab:compileatk} shows compilation success rates for the subset of 34 natural Dart functions. Decompiler-v1 achieves 79.4\% success at $k{=}5$, a 14.7 percentage point improvement over the base model. However, the wide confidence intervals (63.2--89.7 for v1, 47.9--78.5 for base) indicate this difference is not statistically significant ($p \approx 0.18$). The single-attempt performance (\compileatk$_1$) shows similar trends with overlapping confidence intervals.

\begin{table}[!t]
\centering
\caption{Executability as \compileatk{} for Dart (34 natural functions from real codebases). Two-proportion $z$-test (v1 vs.\ base): at $k{=}5$, $p \approx 0.18$ (n.s.); at $k{=}1$, $p \approx 0.63$ (n.s.).}
\label{tab:compileatk}
\begingroup
\resizebox{\columnwidth}{!}{%
\begin{tabular}{@{}lcc@{}}
\toprule
\textbf{Model} & \textbf{\compileatk$_5$ (95\% CI)} & \textbf{\compileatk$_1$ (95\% CI)} \\
\midrule
Qwen3-4B (base) & 64.7\% (47.9--78.5) & 44.1\% (28.9--60.5) \\
Decompiler-v1 (4B Dart) & 79.4\% (63.2--89.7) & 50.0\% (34.1--65.9) \\
Decompiler-v2 (4B +Swift) & 64.7\% (47.9--78.5) & 38.2\% (24.0--55.0) \\
\bottomrule
\end{tabular}
}%
\endgroup
\end{table}

\subsection{Compiled-Only \codebleu{} Analysis}
Table~\ref{tab:compiledonly} examines \codebleu{} scores only for successfully compiled outputs, revealing an important trade-off. While Decompiler-v1 achieves higher compilation rates, its mean \codebleu{} on compiled outputs (0.5173) is lower than the base model (0.6074). This suggests that the specialized model successfully tackles harder decompilation problems that the base model cannot compile, even if the resulting code has lower similarity to references.

\begin{table}[!t]
\centering
\caption{\codebleu{} on successfully compiled outputs only (Dart natural functions; normalized 0--1 scale), with 95\% CIs for the mean.}
\label{tab:compiledonly}
\begingroup
\resizebox{\columnwidth}{!}{%
\begin{tabular}{@{}lccccc@{}}
\toprule
\textbf{Model} & \textbf{Samples} & \textbf{Avg.} & \textbf{SD} & \textbf{95\% CI} & \textbf{Min--Max} \\
\midrule
Qwen3-4B (base) & 22 & 0.6074 & 0.1416 & 0.5482--0.6666 & 0.3126--0.8814 \\
Decompiler-v1 (4B Dart) & 27 & 0.5173 & 0.1542 & 0.4591--0.5755 & 0.0067--0.8390 \\
Decompiler-v2 (4B +Swift) & 22 & 0.5037 & 0.1179 & 0.4544--0.5530 & 0.2099--0.7655 \\
\bottomrule
\end{tabular}
}%
\endgroup
\end{table}

This inverse relationship between coverage and similarity scores illustrates the challenge in decompilation: models sometimes struggle with certain decompilation tasks, in the base models the output didn't function at all while after optimization it was nudged till it compiled.

\subsection{Quality Analysis (Idiomaticity)}
A preliminary human evaluation (single reviewer) indicated that specialized models produce more idiomatic Dart code with meaningful variable names and appropriate use of language-specific constructs (e.g., null-safety operators, arrow functions). However, without inter-rater reliability metrics, these observations remain anecdotal and require future systematic evaluation.

\section{Discussion: Synthetic vs.\ Close-Language Training}

\subsection{Observed Patterns in Multi-Language Training}
Our experiments reveal scale-dependent effects when adding Swift to Dart training data. At 4B parameters, Dart+Swift (v2) has lower Dart performance by approximately 2.1 \codebleu{} points, while at 8B, it improves performance by 8.7 points. This divergence suggests fundamental differences in how models of different capacities handle multi-language learning.

\subsection{Capacity, Interference, and Representational Bottlenecks}
We hypothesize four interacting factors explain the scale-dependent behavior:

\textbf{Capacity saturation.} Performance improves with model size, dataset size, and compute, with larger models being more sample-efficient~\cite{scalinglaws2020}. At 4B parameters, the model lacks sufficient capacity to effectively encode patterns from both languages, leading to competition for limited parameters.

\textbf{Cross-lingual interference.} Multilingual training can induce negative transfer when capacity is insufficient to disentangle languages or build language-agnostic abstractions. While Dart and Swift share some common idioms (optional types, closures), they differ in crucial aspects (memory management, syntax). At 4B scale, these differences create confusion; at 8B scale, the model can leverage commonalities while maintaining language-specific representations~\cite{muennighoff2023}.

\textbf{Representational bottlenecks.} Decompilation requires aligning low-level assembly patterns to high-level language constructs. With limited capacity, a single representation must serve two related but distinct targets, creating bottlenecks that force harmful parameter sharing across language-specific features.

\textbf{Optimization level mismatch.} An additional confounding factor is that our Dart training data uses AOT-optimized binaries while Swift uses \texttt{-O0} unoptimized code. This mismatch means the assembly patterns differ not only due to language semantics but also due to optimization strategies. At 4B scale, this additional variation may increase interference; at 8B scale, the model may have sufficient capacity to learn optimization-invariant patterns.

\textit{Practical implication:} Cross-lingual training in code decompilation exhibits a capacity threshold: below it (near 4B in our experiments), interference dominates and performance degrades; above it (near 8B), shared structure can be exploited effectively for improved performance.

\section{Limitations}
\label{sec:limitations}
Our study has several important limitations that should guide interpretation of results:

(1) \textbf{Compilation $\neq$ semantic correctness}: We report \compileatk{} as a syntax validation metric but cannot claim semantic equivalence without comprehensive unit tests. Future work should develop test suites for \passatk{} evaluation.

(2) \textbf{Optimization level mismatch in training}: Our Dart training data uses AOT-optimized binaries while Swift training data uses \texttt{-O0} unoptimized code. This introduces an additional confounding variable when comparing Dart-only vs.\ Dart+Swift training, as differences may arise from optimization patterns rather than purely linguistic transfer. Future work should use matched optimization levels across languages.

(3) \textbf{Limited human evaluation}: Idiomaticity assessment relies on a single rater without inter-rater reliability. A systematic human study with multiple annotators and formal rubrics is necessary.

(4) \textbf{Swift optimization levels unexplored}: While we successfully decompile Dart from production-optimized (AOT) binaries, we have not yet evaluated Swift decompilation at higher optimization levels (\texttt{-O2}, \texttt{-O3}). Cross-language transfer effects may differ when both languages use comparable optimization strategies.

(5) \textbf{Training reproducibility}: While we report major hyperparameters, some details (random seeds, warmup schedules) were not systematically logged as the defaults were chosen, potentially affecting reproducibility.

\section{Future Work}
Several promising directions emerge from this study:

\textbf{Optimization-matched training} using Swift compiled at \texttt{-O2} or \texttt{-O3} would eliminate the optimization level confound and test whether cross-lingual transfer improves when both languages exhibit similar assembly patterns.

\textbf{Comprehensive test harness development} for \passatk$_1$/\passatk$_5$ evaluation would enable semantic correctness assessment beyond syntax validation.

\textbf{Robustness studies} across Dart compilation modes (JIT vs.\ AOT, different optimization flags) and instruction set architectures (x86-64, ARM, RISC-V) would test generalization.

\textbf{Systematic human evaluation} with formal rubrics, multiple annotators, and inter-rater agreement ($\kappa$) would quantify idiomaticity improvements.

\textbf{Investigation of idiom transfer mechanisms} between Dart and Swift, identifying which shared patterns help and which cause interference, could inform better training strategies.

\section{Conclusion}
Under realistic compute and data constraints, small specialized LLMs can serve as practical, idiomatic decompilers for modern languages like Dart. In a controlled evaluation setting (temperature 0.2, top-$p$ 0.99, beam 1), a 4B parameter model fine-tuned on assembly-to-Dart pairs approaches the performance of models 120× larger on \codebleu{} metrics and shows improved compilation rates on natural functions—though statistical significance remains limited by sample size. Importantly, these results are achieved on \textbf{production-optimized AOT-compiled Dart binaries}, demonstrating practical applicability to real-world reverse engineering scenarios.

The effectiveness of cross-language training (adding Swift to improve Dart decompilation) depends critically on model capacity, with benefits emerging only above a threshold (approximately 8B parameters in our experiments). Below this threshold, interference between languages degrades performance; above it, shared structures can be effectively leveraged. However, the optimization level mismatch between Dart (AOT) and Swift (\texttt{-O0}) training data introduces an additional confounding factor that requires further investigation.

These findings suggest that specialized small models offer a computationally efficient path toward practical decompilation of modern languages, achieving meaningful improvements with just 1.5 hours of fine-tuning on a single GPU. As the software ecosystem increasingly relies on languages like Dart and Swift, developing effective decompilation techniques for these targets becomes essential for security analysis, legacy code recovery, and program understanding tasks.

\section*{Data Availability}
The complete dataset (1,194 Dart-only and 1,000 Dart+Swift assembly$\rightarrow$code pairs), evaluation harness (\codebleu{} + \compileatk), and training configurations are available at \url{https://github.com/raafatabualazm/MSc-Code}.

\balance

\appendix
\section*{Appendix A: Statistical Details}
\textbf{Wilson intervals (proportions).} For v1 \compileatk$_5$: $27/34 \Rightarrow 63.2$--$89.7$; base: $22/34 \Rightarrow 47.9$--$78.5$. For \compileatk$_1$: v1 $17/34 \Rightarrow 34.1$--$65.9$; base $15/34 \Rightarrow 28.9$--$60.5$. Two-proportion $z$-tests: at $k{=}5$, $p \approx 0.18$; at $k{=}1$, $p \approx 0.63$.

\textbf{Normal CIs (means).} For compiled-only \codebleu{} in Table~\ref{tab:compiledonly}: base ($n{=}22$) $0.6074 \pm 0.0592 \Rightarrow 0.5482$--$0.6666$; v1 ($n{=}27$) $0.5173 \pm 0.0582 \Rightarrow 0.4591$--$0.5755$; v2 ($n{=}22$) $0.5037 \pm 0.0493 \Rightarrow 0.4544$--$0.5530$.

\end{document}